# Graphene and its derivatives for Analytical Lab on Chip platforms

Joydip Sengupta [a], Chaudhery Mustansar Hussain [b, *]

[a] *Department of Electronic Science, Jogesh Chandra Chaudhuri College (Affiliated to University of Calcutta), Kolkata, 700 033, W.B, India*
[b] *Department of Chemistry and Environmental Science, New Jersey Institute of Technology, Newark, NJ, USA*

*Keywords:*
Graphene
Lab-on-chip
Microchannels
Chamber structures
Commercial challenges sustainable perceptions

## ABSTRACT

Since last decade, graphene has materialized itself as one of the phenomenal materials to modern researchers because of its remarkable thermal, optical, electronic, and mechanical properties. Graphene holds enormous potentials for lab on chip (LOC) devices and can provide diverse fabrication routes and structural features due to their special electronic and electrochemical properties. A LOC device can manipulate fluids using microchannels and chamber structures, to accomplish fast, highly sensitive and inexpensive analysis with high yield. Hence, the graphene based LOC devices can constitute a well-controlled microenvironment for both advanced chemical/biological evaluation and low-cost point-of-care analysis etc. This review critically debates the graphene as a prime candidate for microfluidic devices and their future applicability towards various practical applications. Finally, the opportunities and challenges for the future of graphene with respect to their commercial challenges and sustainability perspectives are discussed.

## 1. Modern era of graphene for LOC

"Miniaturization" has become an important aspect of electronic devices which downsized it initially to microelectronics and now to nanoelectronics technology in order to fabricate devices for a wide range of applications. The miniaturization started in the late 1950s with the advent of "planner" process to integrate maximum circuit components within a small area and moved forward with the incorporation of microelectromechanical systems (MEMS) and miniaturized total chemical analysis systems (μTAS). The term "Lab on Chip" was probably first coined by Moser et al. [1] to describe the miniaturized thin film glutamate and glutamine biosensors developed by them. Later on, the term LOC represents a miniaturized device which is capable to scale-down, integrate and automate the conventional single or multiple laboratory operations into a system that fits on a chip with size ranges from millimeters to a few square centimeters.

The LOC employs microfluidics technology for the fabrication, operation and final outcome of the device. Mark et al. [2] made an extensive review regarding the different aspects of LOC including the required characteristics that a material should possess in order to be used in LOC. Later on, Pumera [3] emphasized on the benefits of using nanomaterials in microfluidics and predicted that graphene will be an important part of LOC in near future. In subsequent time, a wide variety of nanomaterials were used in LOC to offer excellent improvement in the end result for many applications [4]. Among various nanomaterials, graphene is one of the top priority materials for the fabrication of LOC device because of its novel structure and exotic characteristics. Super-thin (0.35 nm) and ultra-light honeycomb structure of graphene with a planar density of 0.77 mg/m$^2$ has attracted great interest because of its unique nanostructure with fascinating chemical and physical properties [5]. One atomic thickness of graphene is advantageous for its transparency as a single layer of graphene is nearly 98% transparent to visible light. Graphene crystal structure is the strongest among all known materials. Graphene has outstanding mechanical properties with Young's modulus 1.0 TPa and intrinsic strength of 130 GPa, respectively, while it's breaking strength is 42 N/m. Graphene possess extremely high thermal conductivity with value up to 8000 W/m.K. The ultrahigh room-temperature electron mobility of $2 \times 10^5$ cm$^2$V$^{-1}$s$^{-1}$ established graphene as the most conductive material till now at room temperature, with a conductivity of $1.42 \times 10^6$ S/m and a sheet resistance of 125 Ω/sq. Moreover, graphene suits well with basic technological functions of microfluidics and thus extensively incorporated in LOC devices. The graphene and its derivatives i.e. graphene oxide (GO), reduced graphene oxide (rGO) and even functionalized graphene are extensively used for the fabrication of analytical microfluidic chip. There are some good reviews regarding the graphene based very specialized microfluidic devices such as electrochemical immune devices [6], biosensors [7,8] and for surface-enhanced Raman scattering sensing [9]. However, there is a need of reviewing the application regime of graphene as a whole in LOC in a much broader context to acquire an idea about the future prospect of graphene in LOC technology and this reviews aiming at that.

\* Corresponding author.
 *E-mail address:* chaudhery.m.hussain@njit.edu (C.M. Hussain).

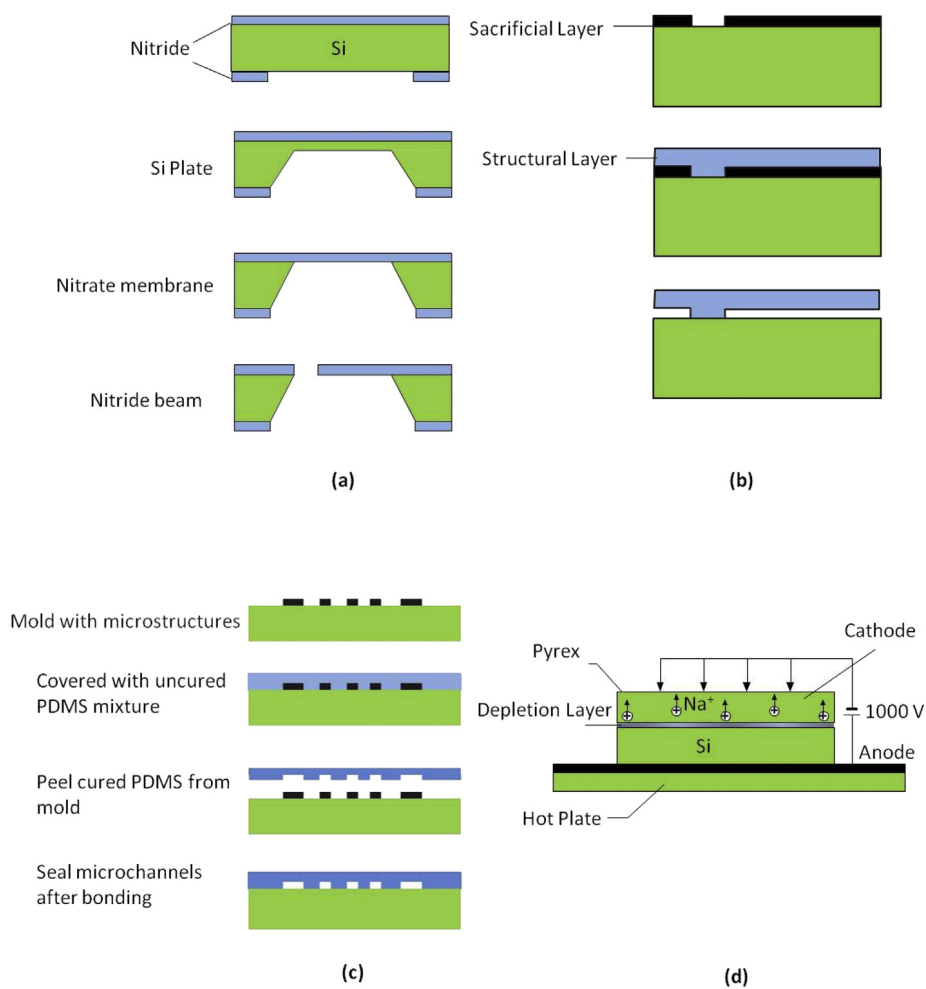

Fig. 1. Different routes LOC fabrication: a) Bulk Micromachining, b) Surface Micromachining, c) Soft Lithography, d) Bonding.

## 2. Lab-on-a-chip miniaturized devices

The commonly used techniques for the fabrication of LOC devices can be divided into two major groups namely top-down and bottom-up. Top-down processes are based on initial patterning on large scale and afterwards reduction of lateral dimensions down to nanoscale. Top-down methods are further classified into three categories: (a) bulk-/film-machining; (b) surface-machining; (c) mold-machining. Bottom-up processes arrange atoms or molecules to create nanostructures by precisely controlled chemical reactions. In the microfluidics based LOC devices fluid transport occur either via laminar co-flow of miscible streams or via segmented-flow of immiscible streams within small channels. Under the main fabrication processes there are many sub processes which are solely or collectively used to achieve the final form of the LOC device. A brief description of the different fabrication subprocesses are delineated here followed by the materials used therein (Fig. 1).

### 2.1. Fabrication processes and subprocesses

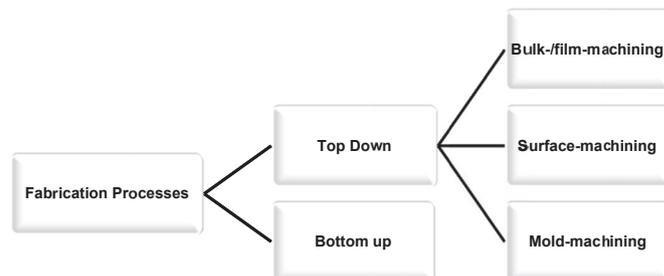

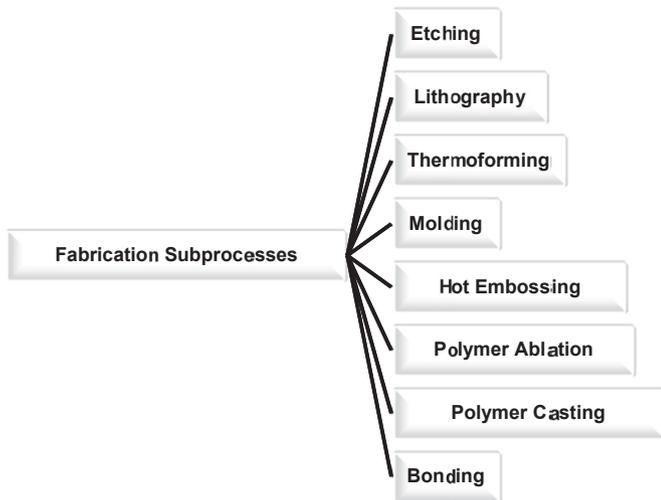

## 2.1.1. Etching

In this process selective part of the material is removed keeping rest of the part intact. The etching is generally used for silicon or glass substrates. The etching process can be subdivided in two categories namely dry and wet etching based on the physical state of the material used to remove. In wet etching liquid chemicals are used as etchants and the nature of the etching process can be isotropic (etching without directional preference) or anisotropic (etching with directional preference). In case of dry etching which is anisotropic in nature, the material removal is performed either by a physical method (molecules with high kinetic energy is used) or chemical method (reactive ion etch (RIE) or inductively coupled plasma (ICP) etch).

## 2.1.2. Lithography

In this process the resist coated substrate is irradiated through a mask to transfer the geometric shapes on a mask onto the surface of the substrate. The exposed resist areas are either soluble or insoluble in a particular solvent named developer. If the irradiated areas are dissolved by the developer then a positive image of the mask is produced and the resist material is called positive resist. When the non-irradiated regions are soluble in the developer then a negative image of the mask appears and the resist is called a negative resist. This principle is identical for all lithography process however the source of irradiation differs, such as UV light, electron beam, X-ray and ion beam. Thus, the related resists and lithography methods are also named after the source.

## 2.1.3. Thermoforming

In this method a thermoplastic sheet is heated till its softening temperature and then pressed against the contours of a mold and later on it cools down and solidifies into the desired shape. The process of thermoforming can be subdivided into six sub process namely drape forming, straight vacuum forming, plug-assist forming, vacuum snap-back forming, pressure-bubble plug-assist vacuum forming, and mechanical forming.

## 2.1.4. Molding

In this method of manufacturing the shape of a pliable material is tailored by employing a rigid frame called mold. There are various processes by which molding can be performed such as injection molding, sintering, transfer molding, reaction injection molding, compression molding, and resin transfer molding. However for the microfluidic applications the most widely used molding process is injection molding.

## 2.1.5. Hot embossing

It is the process of stamping of a pattern using high pressure onto a polymer which has been softened by heating the polymer up to its the glass transition temperature. The stamp can be made by different manner with the most popular being the micro-machining of silicon which is used to govern the pattern in the polymer.

## 2.1.6. Polymer ablation

Microstructures in polymer can also be obtained using ablation of polymer via conventional mechanical drilling, sawing, laser machine and powder blasting. Among these methods laser ablation is preferred due to its higher efficiency over others. Different kinds of laser sources such as excimer laser, $CO_2$ laser, UV laser etc. are used for the ablation of polymer.

## 2.1.7. Polymer casting

In this process the negative of the required design is created with a hard material called mold and then liquid polymer is poured on it. The polymer is cured employing heat or UV radiation and finally peeled off from the mold. Even polymer membrane can be casted directly for the fabrication of microfluidic device.

## 2.1.8. Bonding

To create enclosed fluid channels for fluid handling, the base plate and the cover plate must be bonded after patterning all features on substrates. Different methods are used to achieve a good quality bonding such as silicon direct bonding, glass bonding, plasma-activated bonding, anodic bonding (for silicon or glass plates), thermal bonding (for polymer or glass plates), Glass Frit Bonding, reactive bonding and adhesive bonding, fusion bonding, polymer, and eutectic bonding.

## 2.2. Materials for fabrication

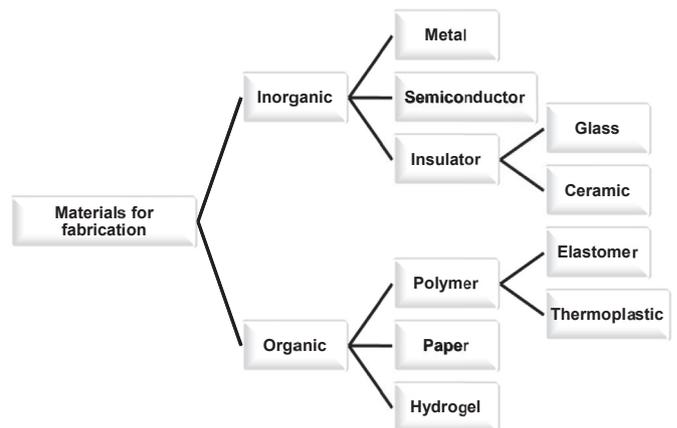

Different kinds of materials are used to manufacture microfluidic devices and they can be broadly categorised as inorganic, organic and composite materials.

### 2.2.1. Inorganic materials

At the infant stage of LOC, the semiconducting material Si was used as a platform for the production of microfluidic device due to its good mechanical properties, suitability for etching, ability of structuring, medical compatibility and chemically stability. However production cost of Si is on the higher side. Glass which is an insulator by nature is also used as building material for microfluidic devices as it is economical, optically transparent, and biologically adaptable with low thermal expansion coefficient. For the fabrication of LOC devices low-temperature cofired ceramic (LTCC) [10] is also used. LTCC is produced by firing of the low temperature (less than 1000°C) sintered multilayer ceramic with the high conductivity thick-film electrodes (e.g. Ag or Cu), simultaneously. Three dimensional structures with numerous layers can be fabricated using LTCC which offers a high level integration of embedded devices, electronic circuits, and fluid networks. Metal in its liquid from is also advantageous for the formation of miniaturized microfluidic modules including heaters, pumps, valves, and electrodes.

#### 2.2.2. Organic materials

Different kinds of organic materials are used for the fabrication of LOC devices such as polymer, paper and hydrogel. In polymer category mainly two kinds of polymers likewise elastomers [Polydimethylsiloxane (PDMS), thermoset polyester (TPE)], and thermoplastic [polystyrene (PS), polycarbonate (PC), poly-methyl methacrylate (PMMA), poly-ethylene glycol diacrylate (PEGDA), polyurethane (PU)] are used. A new class of polymer namely cyclic olefin copolymer(COC) is also used for the manufacturing of LOC device. In recent times, paper the cellulose-based, flexible material has become a prospective microfluidic substrate as it is readily available, biocompatible and inexpensive. However the poor mechanical properties restrict the extensive use of the paper as a substrate material. Hydrogel is a macromolecular polymer gel made of cross-linked polymeric network and demonstrates the capability to swell and absorb a significant portion of water within its structure, but not dissolvable in water. Nowadays it is used for the fabrication of LOC devices especially in biological research however the bonding of hydrogel remains a critical issue for fabrication.

A detailed process flow for the fabrication of microfluidic device can be found in the article by Lake et al. [11].

### 3. Graphene

#### 3.1. Synthesis techniques of graphene

Graphene can be synthesised using two approaches namely top down and bottom up (Fig. 2).

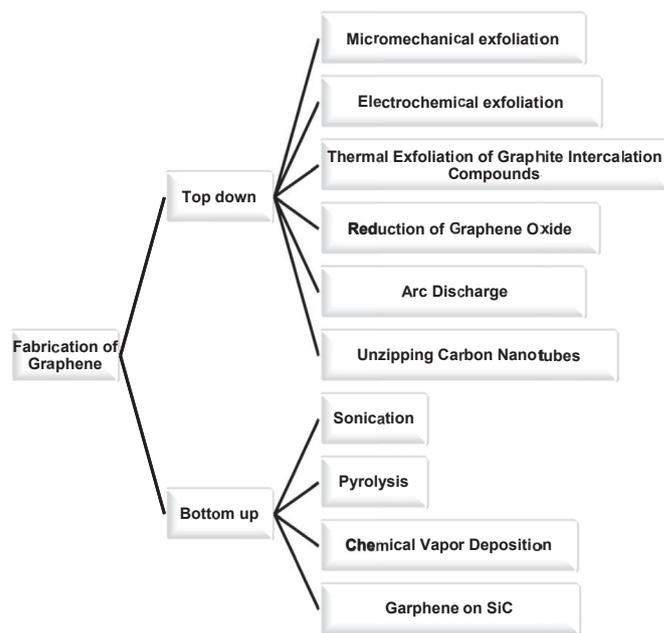

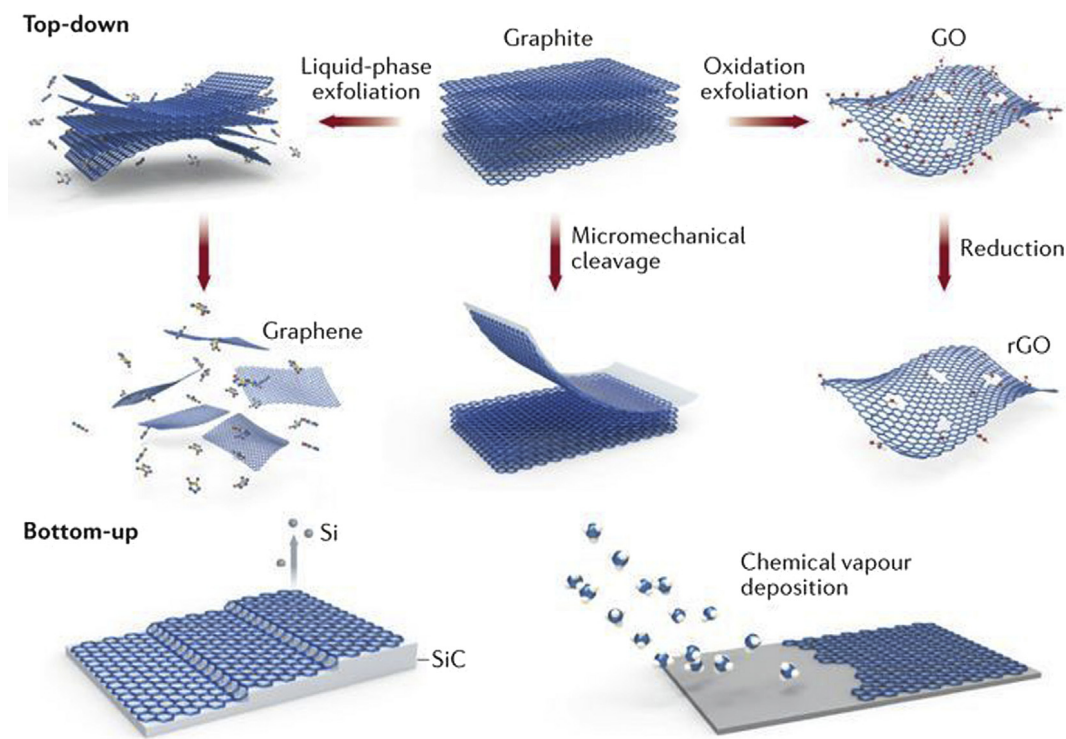

**Fig. 2.** Different routes of graphene synthesis. (Reprinted with permission from ref [96]. Copyright 2017 Springer Nature).

### 3.1.1. Top down approach

This approach involves the methods to produce single graphene sheet by disassembling the stacked layers of graphite.

*3.1.1.1. Micromechanical exfoliation.* In this method adhesive tapes (e.g. scotch tape) is used to peel off the layers from graphite and can be used to synthesise mono-, bi- and few-layer graphene. This is the first method [12] which was used to experimentally isolate graphene from graphite. The resultant graphene are of high quality however the method is time consuming and laborious. Thus the graphene produced using micromechanical exfoliation is generally used to study the fundamental properties instead of commercial applications.

*3.1.1.2. Electrochemical exfoliation.* At the initial stage graphite was used as a sacrificial electrode in the electrochemical set-up for the exfoliation of graphite, and the exfoliated material was collected from the electrolyte solution. With the advancement of the techniques, nowadays surfactant (to prevent reagglomeration of the graphene) and $H_2SO_4$–KOH solutions ($H_2SO_4$ is a strong oxidation of graphite and KOH lowers the acidity of the electrolyte solution) are mixed within electrolytes. In general a mixture of graphite flakes with varying thickness can be produced by electrochemical exfoliation and few-layer graphene is isolated by centrifugation.

*3.1.1.3. Thermal exfoliation of graphite intercalation compounds.* In this process graphite is reduced to graphene via intercalation, on application of thermal energy. Various chemical species can be inserted between graphite interlayer spacing to synthesise graphite intercalation compounds (GIC). In general, firstly, the graphite is exposed to strong acids to yield a GIC and later exposed to rapid thermal heating or to microwave radiation in order to thermally decompose the intercalates into gaseous species which breaks the layers apart. Different intercalates will results in numerous GICs having unique properties to enhance electrical, thermal and magnetic performance.

*3.1.1.4. Reduction of graphene oxide.* Graphite oxide is exfoliated more easily than graphite and this phenomenon is utilized to produce graphene from graphite oxide via chemical reduction. The synthesis of GO is generally performed via the oxidation of graphite employing oxidants and concentrated acids via Hummers [13], Brodie [14] or Staudenmaier methods [15]. Afterwards graphite oxide is exfoliated to produce GO which is later reduced to graphene. However, the end material is usually termed as 'reduced graphene oxide' (rGO) instead of 'graphene' as complete reduction is yet to be achieved. Exfoliation of graphite oxide is performed employing thermal treatments or through sonication in water. The resulting GO is reduced afterwards via thermal or chemical methods to obtain graphene.

*3.1.1.5. Arc discharge.* In this method a DC current is passed in-between graphite electrodes of high purity in presence of high pressure of hydrogen gas or hydrogen-inert gas mixture to synthesise few-layer graphene. Shen et al. [16] found that a mixture of helium and hydrogen gas produces the highest crystallinity material among different gases studied. The growth mechanism involves evaporation of graphite and reactive-gas-confining crystallization of the evaporated carbon molecules. An aqueous arc discharge process was reported by Kim et al. [17] to produce bi- and tri layer graphene which was exfoliated from the graphite electrodes via thermal expansion along with water cavitations employing rapid heating. Controllable stacked number of graphene layers was obtained with reduced oxygen-related defects by adjusting the arc discharge power.

*3.1.1.6. Unzipping carbon nanotubes.* Single or multi-walled carbon nanotubes can be unzipped using physical (such as laser irradiation, electric field and plasma etching) or chemical method (using wet chemistry methods) to produce graphene or few-layer graphene, respectively. This unzipping results in a thin elongated strip of graphene called graphene 'nanoribbons', with ribbon widths governed by the tube diameter. Fission via C–C bond of carbon nanotube is solely responsible for the unzipping, which is often initiated at defect sites, leading to irregular edges [18]. Later on, the growth of nanoribbons having well regulated edges was synthesised [19] by unzipping of flattened carbon nanotubes, where attack preferentially occurs along the bent edges.

*3.1.1.7. Sonication.* In this method, ultrasonic energy is used for the exfoliation of unmodified, natural flake graphite to separate graphene flakes dispersed in solvents. Though different types of ionic solvents are used till date, but the experimental evidence revealed that NMP (N-methyl-pyrrolidone) is the best solvent with respect to the dispersion percentage of monolayer graphene, however cyclopentanone offers the highest absolute concentration (mono- and few-layer graphene). In general, concentration of graphene in solution can be increased by longer sonication time. Sonication can also be performed in aqueous surfactant solutions, which allows avoiding the use of expensive solvents while the surfactants prevent re-aggregation of the graphene.

### 3.1.2. Bottom-up methods

These methods involve preparation of larger graphene using alternative carbon containing sources of smaller dimensions.

*3.1.2.1. Pyrolysis.* In this process graphene is synthesised by low-temperature flash pyrolysis of a solvothermal product of sodium and ethanol. Initially the graphene was obtained as fused sheets, which is separated out into individual sheets by mild sonication in ethanol. The advantage of this process was scalable low-cost fabrication of high-purity graphene at low temperature using non-graphitic precursors. However, the quality of graphene was not up to the mark as it consists of a large number of defects.

*3.1.2.2. Chemical vapor deposition.* Chemical vapor deposition (CVD) can be used to grow graphene and few-layer graphene (FLG) by decomposition of hydrocarbon gases on catalytic surfaces or by surface segregation of C dissolved within the bulk of specific metals. Solubility limit of carbon in the metal is responsible to decide whether the former or the latter will be the dominant growth process. Though various metal substrates like Ni, Fe, Ru, Co, Rh, Ir, Pd, Pt, Cu, Au and semiconductor like Ge even insulator [20] were used for graphene growth, but till date Cu and Ni remains as the most popular. The polycrystalline films of both Cu and Ni can be used for continuous graphene growth. The solubility of C in Cu is less than 0.001 atom% at 1000°C while in Ni the solubility of C is 1.3 atom% at 1000°C. The difference in solubility limits resulted in various growth mechanism of graphene for Cu and Ni. Surface catalyzed mechanism is accountable for the graphene growth on Cu, resulting in monolayer graphene growth under UHV conditions. On the other hand, segregation growth mechanism is responsible for the graphene growth on Ni where UHV conditions are omitted but the process is difficult to control. Substrate free growth is also reported [21] in case of graphene and this method yield large quantities of graphene.

*3.1.2.3. Epitaxial growth on silicon carbide.* When SiC crystal (3C–SiC, 4H–SiC or 6H–SiC) is subjected to high temperature annealing under vacuum for graphitisation, the top layers of SiC crystal is thermally decomposed resulting in desorption of Si atoms. The C atoms remaining on the surface rearrange themselves and re-bonded to form epitaxial graphene layer. Presence of argon or small quantities of disilane reduces the rate of silicon sublimation, resulting in allowance of higher temperatures to get higher quality graphene. Graphene formation initiates at the top surface of SiC and nearly three Si–C bilayers decompose (~0.75 nm) to form one graphene layer (~0.34 nm) and proceeds inwards further.

### 3.2. Functionalization of graphene

Functionalization of graphene and its derivatives is precisely significant to amend their characteristics and inflating the opportunities in both fundamental studies as well as in different device platforms such as LOC. Graphene possesses exotic physical characteristics; however, it has tendency to agglomerates or even restack that could hinder its potential application especially in microfluidic devices. This phenomenon can be countered through functionalization of graphene as it will create strong polar-polar interactions of hydrophilic groups to avoid agglomeration. Though graphene is considered as chemically inert, however experimentally produced graphene has several defects (e.g. vacancies, edges, curvatures and chemical impurities) which increases the reactivity of the graphene [22]. Using these defects, functionalization of graphene can be achieved either by forming strong and stable covalent bonds or via non-covalent bonding [23] as shown in diagram below.

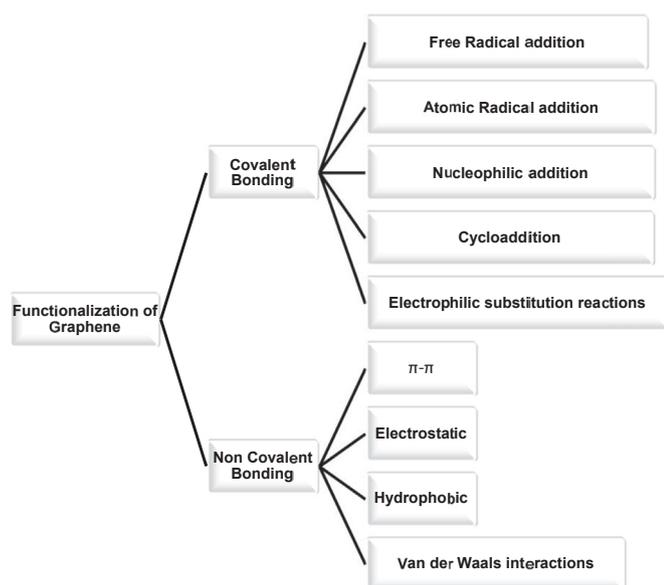

The covalent functionalization increases the dispersibility of graphene in both organic solvents and water. The covalent functionalization can be further classified as free radical addition, atomic radical addition, nucleophilic addition, cycloaddition and electrophilic substitution reactions.

Atoms or groups of atoms having an unpaired electron are called free radicals which are highly reactive and it interacts with the graphene to form covalent bonds via thermal or photochemical treatments. The radicals which are commonly used are aryl diazonium salts, benzoyl peroxide employing the processes like the Bergman cyclization [24] and the Kolbe electrosynthesis [25]. Instead of using organic free radicals if graphene is functionalized using atomic radicals like hydrogen, fluorine and oxygen then more uniform and homogeneous functionalization can be achieved. Covalent modification of graphene can also be accomplished via nucleophilic addition reaction where graphene acts as an electron acceptor. Another way of achieving covalent functionalization of graphene is cycloadditions whereby two π bonded molecules interact and create a new cyclic molecule by forming two σ bonds. Various kinds of cycloadditions can be performed on graphene such as [2 + 1], [2 + 2], [3 + 2] and [4 + 2]. The electron-rich nature of graphene makes it possible to functionalize graphene using electrophilic substitution reactions. The Friedel–Crafts acylation [26] and hydrogen–lithium exchange [27] are the main processes via which electrophilic substitution reactions with graphene can be achieved.

On the other hand, high specific surface area of graphene is advantageous for non-covalent functionalization by physical adsorption of molecules which relies on hydrophobic effects, π-effects, van der Waals forces and electrostatic effects.

## 4. Applications of graphene lab on chip platforms

Though the use of graphene in LOC is still in its infant stage but already a wide variety of devices is fabricated using graphene.

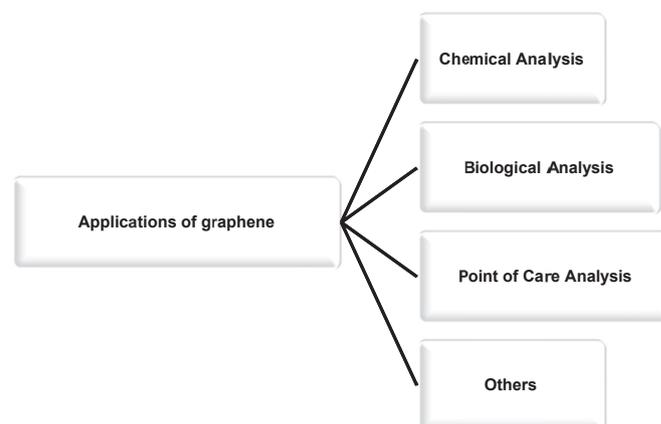

### 4.1. Chemical analysis

Since its birth, Graphene has been widely used for the purpose of electrochemical sensing [28]. Pumera and his coworkers used graphene based nanomaterials for electrochemical detection in LOC devices. However, at initial stage they could not found any benefit of using GO as electrochemical detector in microfluidics device [29]. Later on they found that electrochemically reduced graphene can be employed as a detector which depicted remarkable performance [30]. They also suggested that the detection limits of different nanomaterials (e.g. silver nanoparticles) can be potentially improved by employing graphene-based electrochemical detector as well [31]. Xu et al. [32] designed a paper-based solid-state electrochemiluminescence sensor using functionalized graphene/nafion composite film. To monitor soil ions for nutrient management Ali et al. [33] synthesised bio scaffolds based on graphene foam-titanium nitride to fabricate microfluidic sensor platform. They also used GO nanosheets to design [34] a microfluidic impedimetric nitrate sensor for precise detection and quantification of nitrate ions. For the simultaneous detection and removal of contaminants such as polybrominated diphenyl ethers, Chałupniak et al. [35] developed graphene based miniaturized microfluidic platform (Fig. 3). For the electrochemical detection of heavy metals, GO-polydimethylsiloxane (GO-PDMS) based LOC platform was used by Chałupniak et al. [36]. Zhao et al. [37] fabricated humidity sensor using GO where CMOS interdigital capacitance was used as the basic structure. Park et al. [38] incorporated GO quantum dot into a microfluidic device for trace lead detection. Detection of 4-aminophenolin pharmaceutical paracetamol formulations was achieved by Rattanarat et al. [39] using the graphene-polyaniline modified carbon paste electrode coupled with droplet-based microfluidic sensor. To determine different chemical contaminants in food, Zhang et al. [40] fabricated a paper-based microfluidic device which is integrated with fluorescence labeled single-stranded DNA functionalized GO sensor.

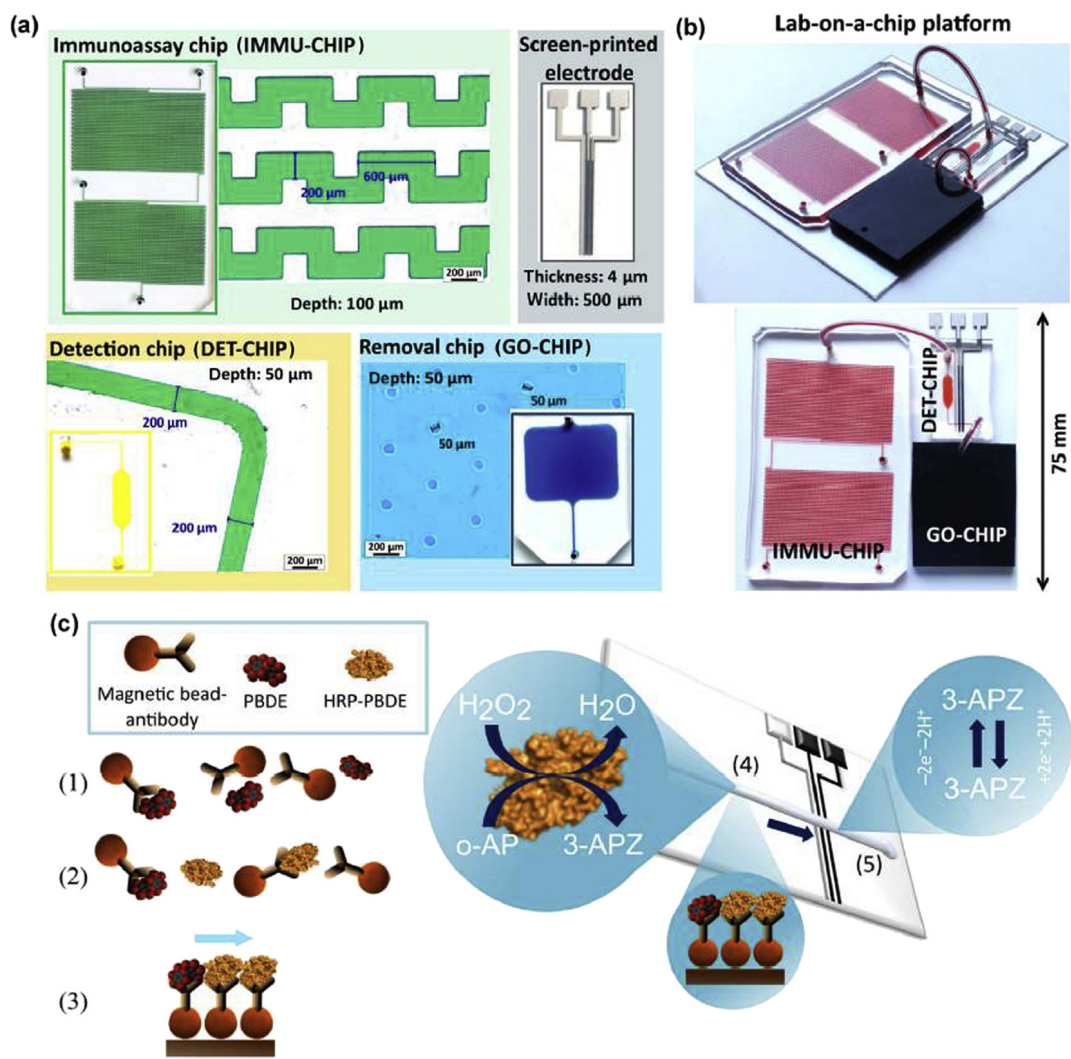

**Fig. 3.** LOC system for detection and removal of PBDEs. (a) Components of the LOC device with visualization of the microfluidic structures (10 × magnification objective); (b) side and top view of the assembled LOC device on a polycarbonate substrate with a screen-printed carbon electrode (SPCE); (c) assay procedure: (1) on-chip mixing and incubation of magnetic beads coupled to anti-PBDE antibodies (MB-Ab) with samples containing PBDE, (2) on-chip mixing and incubation of MB-Ab with HRP-PBDE, (3) immobilization of the immunocomplex in chip using an external magnet and washing of unbound molecules, (4) incubation with substrate solution (o-AP + $H_2O_2$), and (5) electrochemical detection of the product of enzymatic reaction (3-aminophenoxazone (3-APZ)). (Reprinted with permission from ref [35]. Copyright 2016 Springer Nature).

### 4.2. Biological analysis

Graphene based LOC is a promising candidate for biological analysis (Fig. 4) Xiang et al. [41] developed a microscale fluorescence-based colorimetric sensor employing a graphene nanoprobe for multiplexed DNA analysis. A power-free microfluidic chip was developed by Li et al. [42] for fluorescent DNA detection employing a DNA intercalator and GO. For multiplex quantitative LAMP detection for the amplification of DNA Duo et al. [43] fabricated a paper/poly(methyl methacrylate) (PMMA) hybrid CD-like micro-fluidic SpinChip integrated with DNA probe-functionalized GO nanosensor which can deal with the problems faced by conventional mLAMP in identifying multiple targets. A highly flexible microfluidic tactile sensor based on GO nanosuspension was reported by Kenry et al. [44]. A pocket-size, electrochemical microfluidic sensing device was manufactured by Yang et al. [45] using GO nanocomposites for real-time glucose monitoring in human urine samples. Chou et al. [46] developed a $RuO_2$/Graphene/Magnetic Bead-GOx-Nafion based glucose biosensor integrated with microfluidic device to improve sensitivity. Viswanathan et al. [47] devised an approach to develop a graphene-based biosensor via addition of functionalized graphene on a field-effect transistor for glucose detection. Detection of hypoglycemia in clinics is potentially possible with the microfluidic chip devised by Pu et al. [48] where graphene was used to enhance the electroactive nature of the electrode to facilitate precise assessment of low levels of glucose. Zhang et al. [49] developed a facile microfluidic approach to prepare nanostructured GO/Polydopamine coating and used this distinctive biocompatible nano-interface to develop a microfluidic exosome sensing platform. Oh et al. [50] demonstrated that the peptide nucleic-GO complex can be employed to observe exosomes and also for monitoring individual cellular expression of mature microRNAs. To detect different metabolites Labroo et al. [51] designed biosensor arrays based on graphene-ink on a microfluidic paper. Sui et al. [52] used single-layer graphene to manufacture ultra-thin microfluidic devices to perform on-chip protein crystallography analysis using X-ray diffraction with high signal-to-noise ratio. They further improved their LOC device to perform electrocrystallization experiments and concluded that this technology will be useful for wide range of applications of microfluidics [53,54]. Negatively charged GO was used by Bao et al. [55] to fabricate microchip bioreactor for microfluidic proteolysis.

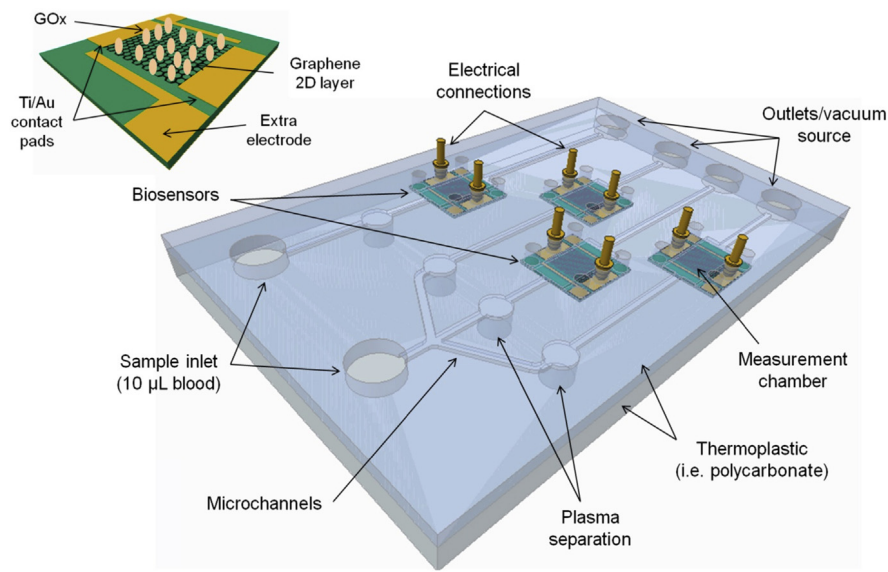

**Fig. 4.** Example of a plastic-based design for a microfluidic platform for multiple, standardized biosensor chips (i.e. graphene FET sensor). (Reprinted with permission from ref [47]. Copyright 2015 Elsevier).

### 4.3. Point of care analysis

Point of care analysis is another area which can be addressed well with the help of graphene based LOC (Fig. 5). Batalla et al. [56] developed a point of care device using polymer/graphene-based electrodes for *in situ* detection of detection of D-amino acids which are associated with serious diseases caused by Vibrio cholerae. To sense malaria-infected red blood cells, a graphene transistor array integrated with microfluidic flow cytometry was developed by Ang et al. [57]. A microfluidic biosensor for peanut allergen detection was reported by Weng et al. [58,59] employing GO and aptamer-functionalized quantum dots. Singh et al. [60] used GO nanocomposite to fabricate microfluidic immunochip to detect of Salmonella typhimurium bacterial cells with high sensitivity and selectivity. For one-step multiplexed pathogen detection Zou et al. [61] designed a PDMS/paper/glass hybrid microfluidic biochip integrated with functionalized GO nano-biosensors. For label-free detection of influenza viruses Singh et al. [62] designed rGO-based electrochemical immunosensor. Chan et al. [63] fabricated a microfluidic device incorporating rGO transistor via a flow-through strategy for the detection of H5N1 influenza virus with high sensitivity and stability. For one-step detection of norovirus Chand et al. [64] fabricated microfluidic platform integrated with graphene-gold nano-composite aptasensor. For ultrasensitive antigen detection Haque et al. [65] used electrochemically rGO-based electrochemical immunosensing platform. An electroluminescence immunosensor based on gold/graphene modified screen-printed working electrode was employed by Yan et al. [66] to develop a microfluidic origami immunodevice for sensitive point-of-care testing of carcinoembryonic antigen. To electrochemically detect carcinoembryonic antigen with high sensitivity Wang et al. [67] used functionalized graphene nanocomposite to fabricate a unique label-free microfluidic paper-based immunosensor. Zhou et al. [68] detected prostate-specific antigen by using magnetic controlled photoelectrochemical sensing system which was designed employing rGO-functionalized $BiFeO_3$ as the photoactive material. A microfluidic chip integrated with a novel GO-based Forster resonance energy transfer biosensor was fabricated by Cao et al. [69] to detect cancer cells *in situ*. Wu et al. [70,71] fabricated a microfluidic point-of-care electrochemical immunodevice and incorporated graphene based signal amplification strategy for multiplexed measurement of cancer biomarkers. To detect a small quantity of cancer cells Xing et al. [72] developed a graphene-based optical refractive index sensor derived from the polarization-dependent absorption characteristics of graphene under total internal reflection. The diagnosis of cancer at early stage is always preferable as it increases the probability of complete recovery from the disease [73]. For early diagnosis of cancer Yoon et al. [74] explored the opportunity of integrating nanotechnology with established cancer research to design a microfluidic device with GO for sensitive and selective detection of circulating tumor cells. Ali et al. [75] fabricated a microfluidic immunosensor to detect breast cancer molecules utilizing a graphene foam modified electrode. A 3D biointerface of graphene-based electrical impedance sensor was fabricated by Wang et al. [76] for the diagnosis of metastatic cancer. Inoue et al. [77] fabricated graphene-based inline pressure sensor integrated with microfluidic elastic tube which can be used for accurate pressure measurement in microfluidic systems including organ-on-a-chip devices. Jiao et al. [78] used graphene, microfluidic liquid metal, and stretchable elastomer to fabricate an all-flexible strain sensors to monitor structural health of curved concrete structures, and also to track the angular motion of human wrist.

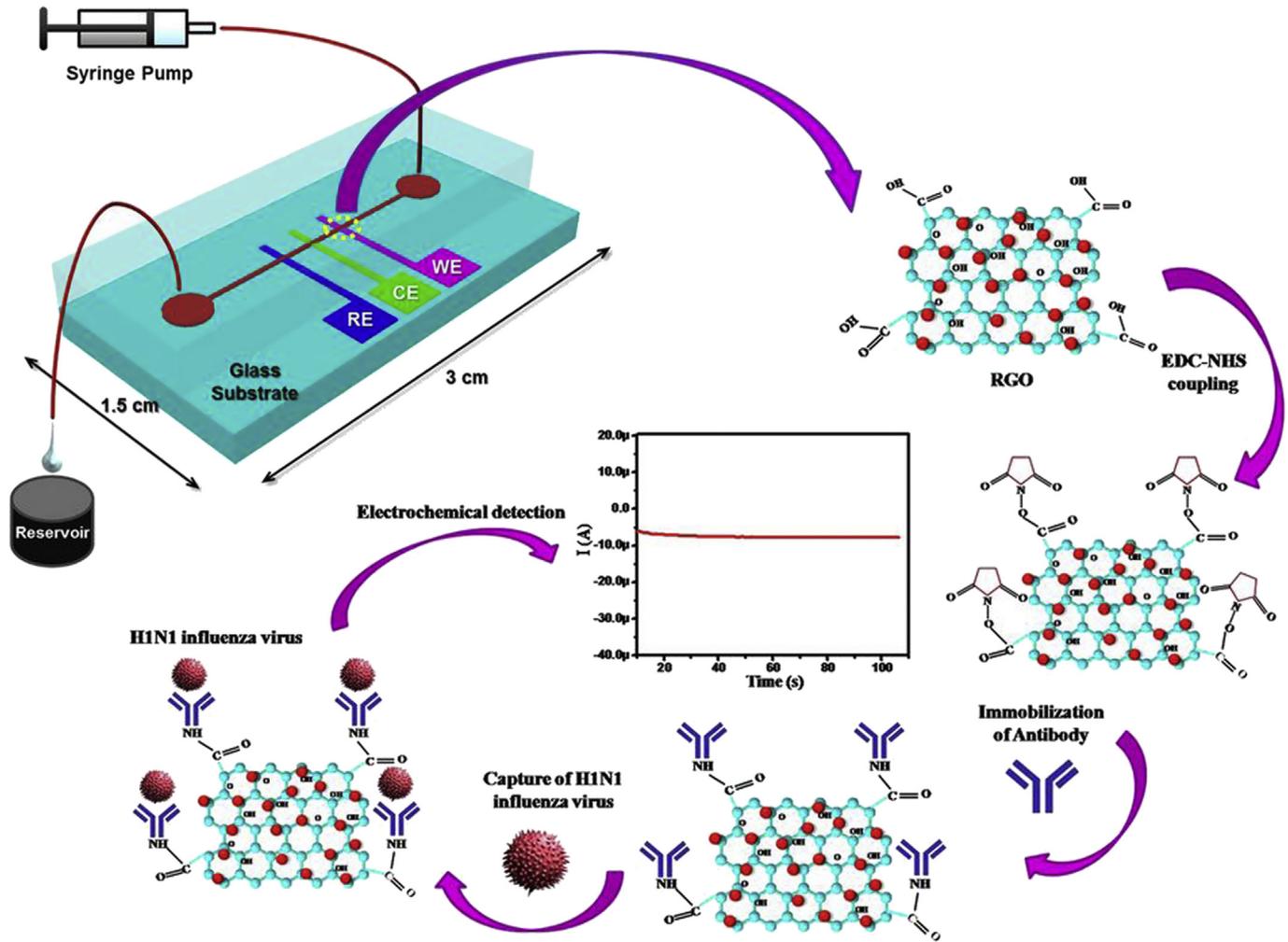

**Fig. 5.** Schematic illustration of the microfluidics-integrated electrochemical immunosensing chip coated with RGO, followed by antibody immobilization using EDC/NHS coupling for the detection of influenza virus H1N1. (Reprinted with permission from ref [62]. Copyright 2017 Springer Nature).

### 4.4. Other applications

In recent past, graphene evolved as a favorable choice for optoelectronic applications specially photodetection [79]. Now the as integration of photo detector in a chip is handful for several application so Gan et al. [80] used mechanically exfoliated bilayer graphene to fabricate a chip-integrated ultrafast graphene photo detector which exhibited large responsivity, high speed and wide spectral bandwidth. The industrial chip design flow is based on CMOS technology so a CMOS-compatible graphene based photo detector is highly appreciable and Pospischil et al. [81] worked in this direction to make CMOS-compatible graphene based photo detector encompassing all optical communication bands. For optical communications by remote transfer of heat Miyoshi et al. [82] fabricated high-speed and on-chip graphene blackbody emitters. For label-free imaging of the electric field dynamics in solutions Horng et al. [83] demonstrated a method that employed the exclusive gate-variable optical transitions of graphene with a critically coupled planar waveguide platform. To probe the electrical signals in mouse retina Zhang et al. [84] combined the graphene field-effect transistors and scanning photocurrent microscopy with microfluidic platforms (Fig. 6).

## 5. Commercial challenges

### 5.1. Economics of graphene

Recently, Zou et al. [85] studied the research trends of graphene in the world by employing the data from Chemical Abstracts Service. They summarized that both papers and patents in graphene has increased in numbers with time which is the representation of continuous growth of graphene based sectors with potential application in diverse fields and also expanding the usage limit. They further concluded that there are still challenges in technological breakthrough regarding its synthesis methods and further processing to achieve graphene industrialization. A study by the famous management consulting firm McKinsey & Company [86] indicated that graphene fabrication must be compatible with existing complementary metal-oxide semiconductor (CMOS) devices in order to be the next S-curve for semiconductors.

### 5.2. Health and safety-toxicity of graphene

Increase in global production quantities and a broadened application spectrum of graphene have raised the human and environmental exposure to graphene based nanomaterials. This in turn augmented the anxiety about the potential adverse effects on of graphene on living organisms [87]. A numerous number of *in vitro* and *in vivo* toxicological investigations both short and long term, have assessed the interactions between graphene-based nanomaterials and diverse living bodies. The different studies found that size (specially lateral), shape, fabrication methods, post-fabrication processing steps, purity, presence of reactive oxygen, attached functional groups, route and dose of administration, and exposure times are the responsible parameters for the toxicity of graphene [88]. The cellular uptake properties of the graphene depends on the size, shape and the morphology while the nature of the functional groups attached with graphene can modify its interactions with living cells. The starting materials and the synthesis methods used for the fabrication of graphene oxide can induce non removable metallic impurities and oxidative wreckage in the final product, which are responsible for toxicity effects. Reactive oxidation species incorporated in graphene oxide causes cell damage while sharp edges of gaphene sheets make direct physical damage of the living cells. To circumvent the toxic nature of graphene some methods like surface passivation, charge manipulation, introduction of biocompatible coatings are proposed in order to achieve healthier and safer environment [89].

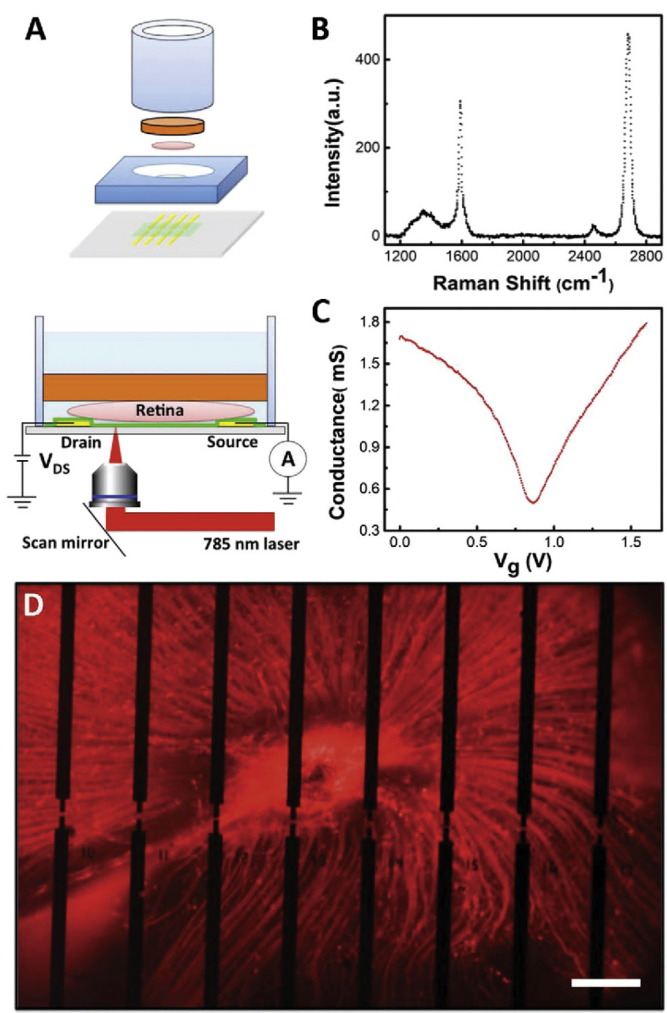

**Fig. 6.** Overview of the experimental design. (A) Top: Exploded view of a graphene-integrated microfluidic platform. The orange plate is a slice of agar gel. The pink disk represents a retina. The green plane indicates a graphene film. The yellow bars represent Au electrodes. Bottom: Schematic diagram of scanning photocurrent measurements. A one micron diameter diffraction-limited laser spot goes through a transparent coverslip to scan over graphene transistors underneath a retinal tissue in a microfluidic platform. (B) Raman spectrum of graphene on a coverslip. (C) Electrolyte gate response of a typical graphene transistor. (D) Fluorescence image of a CTB-labeled retina on top of graphene transistors. There are 54 separated electrodes in an electrode array, where the edge-to-edge distances between two electrodes in the horizontal direction is 180 μm and between upper and lower electrodes are 20 μm, 40 μm, and 60 μm, respectively. A graphene film is placed on top of all electrodes. Any two electrodes can be used as source and drain electrodes, respectively. Here, 16 opaque electrodes near the ONH are imaged. Scale bar is 200 μm. (Reprinted with permission from ref [84]. Copyright 2016 RSC Pub).

### 5.3. Life cycle assessment (LCA)

Arvidsson et al. studied the LCA of graphene synthesised by ultrasonication and chemical reduction [90] and found that the ultrasonication route uses less energy and water; on the other hand chemical reduction route has less impact on human and ecotoxicity. The LCA study of graphene grown by epitaxial method [91] revealed that thinner SiC wafer and use of renewable energy sources has to be implemented in order to achieve the environmental impact. Kahnam et al. [92] compared the LCA of graphene grown by biotechnological process and chemical process and found that in chemical process more energy is consumed (1642 Wh), than in the biotechnological process (5 Wh). All together it can be stated that in each production process of graphene there is a step (reduction step in chemical reduction of graphite oxide, heating step in thermal exfoliation, hydrocarbon gas feedstock for CVD, production of the silicon carbide wafer for epitaxial growth) is mainly responsible for huge environmental impact [93].

## 6. Sustainable perceptions of graphene

Researchers are trying to make graphene and its composite more sustainable in order to achieve a greener world. There are few reports [94,95] about the eco friendly sustainable synthesis of graphene. It can be hoped that more and more paths will be opened up in near future so that a sustainable growth of graphene and its inclusion in LOC can be achieved.

## 7. Conclusions

Graphene based microfluidics combines the advantages of exotic properties graphene with peerless analytical characteristics of microfluidics such as minimal reagents consumption, super fast detection, ease of use along with high sensitivity. This paper firstly introduces the process technology for the fabrication of graphene based LOC and later reviews the major application areas of graphene-based microfluidics which spreads from the detection of chemical food contaminates to electrical signal of retina or even to detection of early stage cancer. The prospective economical perspective of graphene based LOC accelerates the growth of such devices. However issues like toxicity and sustainability are still to be addressed. The reproducibility is of major concern of any nanoscale devices and graphene based LOC falls in it. Since many of graphene based LOCs are used in point of care analysis so the integrity and reproducibility has to undergo extensive quality check which may increase the cost. In summary, although there are still some hurdles to achieve inexpensive, eco-bio friendly, sustainable and reproducible graphene based LOC but it can be hoped that graphene based LOC will surmount the barriers in near future with the help of its enormous potential.

**Acknowledgements**

This work was supported by Science and Engineering Research Board, Department of Science & Technology, India under Project number (SR/FTP/PS-120/2012).

## References

[1] I. Moser, et al., Miniaturized thin film glutamate and glutamine biosensors, Biosens. Bioelectron. 10 (6–7) (1995) 527–532. https://doi.org/10.1016/0956-5663(95)96928-R.
[2] D. Mark, et al., Microfluidic lab-on-a-chip platforms: requirements, characteristics and applications, Chem. Soc. Rev. 39 (2010) 1153–1182. https://doi.org/10.1039/b820557h.
[3] Pumera Martin, Nanomaterials meet microfluidics, Chem. Commun. 47 (20) (2011) 5671–5680. https://doi.org/10.1039/c1cc11060h.
[4] Rüstem Keçili, Sibel Büyüktiryaki, Chaudhery Mustansar Hussain, Advancement in bioanalytical science through nanotechnology past, present and future, Trac. Trends Anal. Chem. 110 (2019) 259–276. https://doi.org/10.1016/j.trac.2018.11.012.
[5] Santosh Bahadur Singh, Chaudhery Mustansar Hussain, Nano-graphene as groundbreaking miracle material: catalytic and commercial perspectives, ChemistrySelect 3 (33) (2018) 9533–9544. https://doi.org/10.1002/slct.201802211.
[6] Mohammad Hasanzadeh, et al., Two dimension (2-D) graphene-based nanomaterials as signal amplification elements in electrochemical microfluidic immune-devices: recent advances, Mater. Sci. Eng. C 68 (2016) 482–493. https://doi.org/10.1016/j.msec.2016.06.023.
[7] Janire Peña-Bahamonde, et al., Recent advances in graphene-based biosensor technology with applications in life sciences, J. Nanobiotechnol. 16 (1) (2018) 1–17. https://doi.org/10.1186/s12951-018-0400-z, 75.
[8] Eden Morales-Narváez, Arben Merkoçi, Graphene oxide as an optical bio-sensing platform: a progress report, Adv. Mater. (2018) 1–12. https://doi.org/10.1002/adma.201805043, 1805043.
[9] Zhuqing Wang, et al., Graphene-based nanoplatforms for surface-enhanced Raman scattering sensing, Analyst 143 (21) (2018) 5074–5089. https://doi.org/10.1039/c8an01266k.
[10] S. Petra, Dittrich, "research highlights, Lab Chip 7 (2007) 663–665. https://doi.org/10.1039/b706143a.
[11] Melinda Lake, et al., Microfluidic device design, fabrication, and testing protocols, Protoc. Exch. (2015) 1–26. https://doi.org/10.1038/protex.2015.069. July.
[12] K.S. Novoselov, et al., Electric field effect in atomically thin carbon films, Science 306 (5696) (2004) 666–669. https://doi.org/10.1126/science.1102896.
[13] W.S. Hummers, et al., Preparation of graphitic oxide, J. Am. Chem. Soc. 80 (1958) 1339. https://doi.org/10.1021/ja01539a017.
[14] B.C. Brodie, On the atomic weight of graphite, Phil. Trans. Roy. Soc. Lond. 149 (1859) 249–259.
[15] L. Staudenmaier, Verfahren Zur Darstellung der graphitsaure, Ber. Dtsch. Chem. Ges. 31 (1898) 1481–1487. https://doi.org/10.1002/cber.18980310237.
[16] B. Shen, et al., Influence of different buffer gases on synthesis of few-layered graphene by arc discharge method, Appl. Surf. Sci. 258 (10) (2012) 4523–4531. https://doi.org/10.1016/j.apsusc.2012.01.019.
[17] S. Kim, et al., Graphene Bi- and trilayers produced by a novel aqueous arc discharge process, Carbon 102 (2016) 339–345. https://doi.org/10.1016/j.carbon.2016.02.049.
[18] S. Cho, et al., Radial followed by longitudinal unzipping of multiwalled carbon nanotubes, Carbon 49 (12) (2011) 3865–3872. https://doi.org/10.1016/j.carbon.2011.05.023.
[19] Y.R. Kang, et al., Precise unzipping of flattened carbon nanotubes to regular graphene nanoribbons by acid cutting along the folded edges, J. Mater. Chem. 22 (32) (2012) 16283–16287. https://doi.org/10.1039/c2jm33385f.
[20] J. Sun, et al., Direct chemical vapor deposition growth of graphene on insulating substrates, ChemNanoMat 2 (1) (2016) 9–18. https://doi.org/10.1002/cnma.201500160.
[21] A. Dato, et al., Substrate-free gas-phase synthesis of graphene sheets, Nano Lett. 8 (7) (2008) 2012–2016. https://doi.org/10.1021/nl8011566.
[22] Liang Yan, et al., Chemistry and physics of a single atomic layer: strategies and challenges for functionalization of graphene and graphene-based materials, Chem. Soc. Rev. 41 (1) (2012) 97–114. https://doi.org/10.1039/c1cs15193b.
[23] Vasilios Georgakilas, et al., "Functionalization of Graphene : covalent and non-covalent approaches , derivatives and applications, Chem. Rev. 112 (2012) 6156–6214. https://doi.org/10.1021/cr3000412.
[24] Xiaowei Ma, et al., Functionalization of pristine graphene with conjugated polymers through diradical addition and propagation, Chem. Asian J. 7 (11) (2012) 2547–2550. https://doi.org/10.1002/asia.201200520.
[25] Santanu Sarkar, Bekyarova Elena, Robert C. Haddon, Reversible grafting of α-naphthylmethyl radicals to epitaxial graphene, Angew. Chem. Int. Ed. 51 (20) (2012) 4901–4904. https://doi.org/10.1002/anie.201201320.
[26] Chua Chun Kiang, Pumera Martin, Friedel-crafts acylation on graphene, Chem. Asian J. 7 (5) (2012) 1009–1012. https://doi.org/10.1002/asia.201200096.
[27] Chengfei Yuan, Wufeng Chen, Lifeng Yan, Amino-grafted graphene as a stable and metal-free solid basic catalyst, J. Mater. Chem. 22 (15) (2012) 7456–7460. https://doi.org/10.1039/c2jm30442b.
[28] Pumera Martin, et al., Graphene for electrochemical sensing and biosensing, Trac. Trends Anal. Chem. 29 (9) (2010) 954–965. https://doi.org/10.1016/j.trac.2010.05.011.
[29] Chua Chun Kiang, Adriano Ambrosi, Pumera Martin, Graphene based nanomaterials as electrochemical detectors in lab-on-a-chip devices, Electrochem. Commun. 13 (5) (2011) 517–519. https://doi.org/10.1016/j.elecom.2011.03.001.
[30] Chua Chun Kiang, Pumera Martin, Chemically modified graphenes as detectors in lab-on-chip device, Electroanalysis 25 (4) (2013) 945–950. https://doi.org/10.1002/elan.201200583.
[31] Chua Chun Kiang, Pumera Martin, Detection of silver nanoparticles on a lab-on-chip platform, Electrophoresis 34 (14) (2013) 2007–2010. https://doi.org/10.1002/elps.201200426.
[32] Yuanhong Xu, et al., Paper-based solid-state electrochemiluminescence sensor using poly(sodium 4-styrenesulfonate) functionalized graphene/nafion composite film, Anal. Chim. Acta 763 (2013) 20–27. https://doi.org/10.1016/j.aca.2012.12.009.
[33] Md Azahar Ali, et al., In situ integration of graphene foam-titanium nitride based bio-scaffolds and microfluidic structures for soil nutrient sensors, Lab Chip 17 (2) (2017) 274–285. https://doi.org/10.1039/C6LC01266C.
[34] Md Azahar Ali, et al., Microfluidic impedimetric sensor for soil nitrate detection using graphene oxide and conductive nanofibers enabled sensing interface, Sensor. Actuator. B Chem. 239 (2017) 1289–1299. https://doi.org/10.1016/j.snb.2016.09.101.
[35] Andrzej Chałupniak, Arben Merkoçi, Toward integrated detection and graphene-based removal of contaminants in a lab-on-a-chip platform, Nano Res. 10 (7) (2017) 2296–2310. https://doi.org/10.1007/s12274-016-1420-3.
[36] Andrzej Chałupniak, Arben Merkoçi, Graphene oxide-poly(dimethylsiloxane)-based lab-on-a-chip platform for heavy-metals preconcentration and electrochemical detection, ACS Appl. Mater. Interfaces 9 (51) (2017) 44766–44775. https://doi.org/10.1021/acsami.7b12368.
[37] Cheng-Long Zhao, Ming Qin, Qing-An Huang, Humidity sensing properties of the sensor based on graphene oxide films with different dispersion concentrations, in: 2011 IEEE SENSORS Proceedings, 2011, pp. 129–132. https://doi.org/10.1109/ICSENS.2011.6126968.
[38] Minsu Park, et al., Combination of a sample pretreatment microfluidic device with a photoluminescent graphene oxide quantum dot sensor for trace lead detection, Anal. Chem. 87 (21) (2015) 10969–10975. https://doi.org/10.1021/acs.analchem.5b02907.
[39] Poomrat Rattanarat, et al., Graphene-polyaniline modified electrochemical droplet-based microfluidic sensor for high-throughput determination of 4-aminophenol, Anal. Chim. Acta 925 (2016) 51–60. https://doi.org/10.1016/j.aca.2016.03.010.
[40] Yali Zhang, Peng Zuo, Bang Ce Ye, A low-cost and simple paper-based microfluidic device for simultaneous multiplex determination of different types of chemical contaminants in food, Biosens. Bioelectron. 68 (2015) 14–19. https://doi.org/10.1016/j.bios.2014.12.042.
[41] Xia Xiang, et al., Droplet-based microscale colorimetric biosensor for multiplexed DNA analysis via a graphene nanoprobe, Anal. Chim. Acta 751 (2012) 155–160. https://doi.org/10.1016/j.aca.2012.09.008.
[42] Li Jing, et al., A power-free microfluidic chip for SNP genotyping using graphene oxide and a DNA intercalating dye, Chem. Commun. 49 (30) (2013) 3125–3127. https://doi.org/10.1039/c3cc40680f.
[43] Maowei Dou, et al., A paper/polymer hybrid CD-like microfluidic SpinChip integrated with DNA-functionalized graphene oxide nanosensors for multiplex qLAMP detection, Chem. Commun. 53 (79) (2017) 10886–10889. https://doi.org/10.1039/c7cc03246c.
[44] Kenry, et al., Highly flexible graphene oxide nanosuspension liquid-based microfluidic tactile sensor, Small 12 (12) (2016) 1593–1604. https://doi.org/10.1002/smll.201502911.
[45] Yang Jiang, et al., Nickel nanoparticle-chitosan-reduced graphene oxide-modified screen-printed electrodes for enzyme-free glucose sensing in portable microfluidic devices, Biosens. Bioelectron. 47 (2013) 530–538. https://doi.org/10.1016/j.bios.2013.03.051.
[46] Jung Chuan Chou, et al., Dynamic and wireless sensing measurements of potentiometric glucose biosensor based on graphene and magnetic beads, IEEE Sens. J. 15 (10) (2015) 5718–5725. https://doi.org/10.1109/JSEN.2015.2449906.
[47] Sowmya Viswanathan, et al., Graphene-protein field effect biosensors: glucose sensing, Mater. Today 18 (9) (2015) 513–522. https://doi.org/10.1016/j.mattod.2015.04.003.
[48] Zhihua Pu, et al., A continuous glucose monitoring device by graphene modified electrochemical sensor in microfluidic system, Biomicrofluidics 10 (1) (2016) 1–11. https://doi.org/10.1063/1.4942437, 011910.
[49] Peng Zhang, He Mei, Yong Zeng, Ultrasensitive microfluidic analysis of circulating exosomes using a nanostructured graphene oxide/polydopamine coating, Lab Chip 16 (16) (2016) 3033–3042. https://doi.org/10.1039/c6lc00279j.
[50] Hyun Jeong Oh, et al., Graphene-oxide quenching-based molecular beacon imaging of exosome-mediated transfer of neurogenic miR-193a on microfluidic platform, Biosens. Bioelectron. 126 (2018) 647–656. https://doi.org/10.1016/j.bios.2018.11.027.
[51] Pratima Labroo, Yue Cui, Graphene nano-ink biosensor arrays on a microfluidic paper for multiplexed detection of metabolites, Anal. Chim. Acta 813 (2014) 90–96. https://doi.org/10.1016/j.aca.2014.01.024.
[52] Shuo Sui, et al., Graphene-based microfluidics for serial crystallography, Lab Chip 16 (16) (2016) 3082–3096. https://doi.org/10.1039/c6lc00451b.
[53] Shuo Sui, et al., A graphene-based microfluidic platform for electrocrystallization and in situ X-ray diffraction, Crystals 8 (2) (2018) 1–12. https://doi.org/10.3390/cryst8020076, 76.

[54] Shuo Sui, L. Sarah, Perry, "microfluidics: from crystallization to serial time-resolved crystallography, Struct. Dyn. 4 (3) (2017) 32202. https://doi.org/10.1063/1.4979640.
[55] Huimin Bao, et al., Immobilization of trypsin in the layer-by-layer coating of graphene oxide and chitosan on in-channel glass fiber for microfluidic proteolysis, Analyst 136 (24) (2011) 5190–5196. https://doi.org/10.1039/c1an15690j.
[56] Pilar Batalla, et al., Enzyme-based microfluidic chip coupled to graphene electrodes for the detection of D-amino acid enantiomer-biomarkers, Anal. Chem. 87 (10) (2015) 5074–5078. https://doi.org/10.1021/acs.analchem.5b00979.
[57] Ang, et al., Flow sensing of single cell by graphene transistor in a microfluidic channel, Nano Lett. 11 (2011) 5240–5246. https://doi.org/10.1021/nl202579k.
[58] Xuan Weng, Neethirajan Suresh, A microfluidic biosensor using graphene oxide and aptamer-functionalized quantum dots for peanut allergen detection, Biosens. Bioelectron. 85 (2016) 649–656. https://doi.org/10.1016/j.bios.2016.05.072.
[59] Xuan Weng, Neethirajan Suresh, Paper-based microfluidic aptasensor for food safety, J. Food Saf. 38 (1) (2018) 1–8. https://doi.org/10.1111/jfs.12412.
[60] Chandan Singh, et al., Biofunctionalized graphene oxide wrapped carbon nanotubes enabled microfluidic immunochip for bacterial cells detection, Sensor. Actuator. B Chem. 255 (2018) 2495–2503. https://doi.org/10.1016/j.snb.2017.09.054.
[61] Peng Zuo, et al., A PDMS/paper/glass hybrid microfluidic biochip integrated with aptamer-functionalized graphene oxide nano-biosensors for one-step multiplexed pathogen detection, Lab Chip 13 (19) (2013) 3921–3928. https://doi.org/10.1039/c3lc50654a.
[62] Renu Singh, Seongkyeol Hong, Jaesung Jang, Label-free detection of influenza viruses using a reduced graphene oxide-based electrochemical immunosensor integrated with a microfluidic platform, Sci. Rep. 7 (2017) 1–11. https://doi.org/10.1038/srep42771. November 2016, 42771.
[63] Chunyu Chan, et al., A microfluidic flow-through chip integrated with reduced graphene oxide transistor for influenza virus gene detection, Sensor. Actuator. B Chem. 251 (2017) 927–933. https://doi.org/10.1016/j.snb.2017.05.147.
[64] Rohit Chand, Neethirajan Suresh, Microfluidic platform integrated with graphene-gold nano-composite aptasensor for one-step detection of norovirus, Biosens. Bioelectron. 98 (2017) 47–53. https://doi.org/10.1016/j.bios.2017.06.026.
[65] Al Monsur Jiaul Haque, et al., An electrochemically reduced graphene oxide-based electrochemical immunosensing platform for ultrasensitive antigen detection, Anal. Chem. 84 (4) (2012) 1871–1878. https://doi.org/10.1021/ac202562v.
[66] Jixian Yan, et al., An origami electrochemiluminescence immunosensor based on gold/graphene for specific, sensitive point-of-care testing of carcinoembryonic antigen, Sensor. Actuator. B Chem. 193 (2014) 247–254. https://doi.org/10.1016/j.snb.2013.11.107.
[67] Yang Wang, et al., A novel label-free microfluidic paper-based immunosensor for highly sensitive electrochemical detection of carcinoembryonic antigen, Biosens. Bioelectron. 83 (2016) 319–326. https://doi.org/10.1016/j.bios.2016.04.062.
[68] Qian Zhou, et al., Reduced graphene oxide/BiFeO3 nanohybrids-based signal-on photoelectrochemical sensing system for prostate-specific antigen detection coupling with magnetic microfluidic device, Biosens. Bioelectron. 101 (2018) 146–152. https://doi.org/10.1016/j.bios.2017.10.027. August 2017.
[69] Lili Cao, et al., Visual and high-throughput detection of cancer cells using a graphene oxide-based FRET aptasensing microfluidic chip, Lab Chip 12 (22) (2012) 4864–4869. https://doi.org/10.1039/c2lc40564d.
[70] Yafeng Wu, et al., Paper-based micro fluidic electrochemical immunodevice integrated with nanobioprobes onto graphene film for ultrasensitive multiplexed detection of cancer biomarkers, Anal. Chem. 85 (2013) 8661–8668. https://doi.org/10.1021/ac401445a.
[71] Yafeng Wu, et al., A paper-based microfluidic electrochemical immunodevice integrated with amplification-by-polymerization for the ultrasensitive multiplexed detection of cancer biomarkers, Biosens. Bioelectron. 52 (2014) 180–187. https://doi.org/10.1016/j.bios.2013.08.039.
[72] Fei Xing, et al., Ultra-sensitive flow sensing of a single cell using graphene-based optical sensors, Nano Lett. 14 (6) (2014) 3563–3569. https://doi.org/10.1021/nl5012036.
[73] Mohammad Hasanzadeh, Nasrin Shadjou, Miguel de la Guardia, Early stage screening of breast cancer using electrochemical biomarker detection, Trac. Trends Anal. Chem. 91 (2017) 67–76. https://doi.org/10.1016/j.trac.2017.04.006.
[74] H.J. Yoon, et al., Nanoassembly of Graphene Oxide for Circulating Tumor Cell Isolation, 2011, pp. 1098–1100. Most.
[75] Md Azahar Ali, et al., Microfluidic immuno-biochip for detection of breast cancer biomarkers using hierarchical composite of porous graphene and titanium dioxide nanofibers, ACS Appl. Mater. Interfaces 8 (32) (2016) 20570–20582. https://doi.org/10.1021/acsami.6b05648.
[76] Xiahua Wang, et al., Three-dimensional graphene biointerface with extremely high sensitivity to single cancer cell monitoring, Biosens. Bioelectron. 105 (2018) 22–28. https://doi.org/10.1016/j.bios.2018.01.012.
[77] Nagisa Inoue, Hiroaki Onoe, Graphene-based inline pressure sensor integrated with microfluidic elastic tube, J. Micromech. Microeng. 28 (1) (2018) 1–7. https://doi.org/10.1088/1361-6439/aa9810, 014001.
[78] Yueyi Jiao, et al., Wearable graphene sensors with microfluidic liquid metal wiring for structural health monitoring and human body motion sensing, IEEE Sens. J. 16 (22) (2016) 7870–7875. https://doi.org/10.1109/JSEN.2016.2608330.
[79] Thomas Mueller, Fengnian Xia, Phaedon Avouris, Graphene photodetectors for high-speed optical communications, Nat. Photon. 4 (5) (2010) 297–301. https://doi.org/10.1038/nphoton.2010.40.
[80] Xuetao Gan, et al., Chip-integrated ultrafast graphene photodetector with high responsivity, Nat. Photon. 7 (11) (2013) 883–887. https://doi.org/10.1038/nphoton.2013.253.
[81] Andreas Pospischil, et al., CMOS-compatible graphene photodetector covering all optical communication bands, Nat. Photon. 7 (11) (2013) 892–896. https://doi.org/10.1038/nphoton.2013.240.
[82] Yusuke Miyoshi, et al., High-speed and on-chip graphene blackbody emitters for optical communications by remote heat transfer, Nat. Commun. 9 (1) (2018). https://doi.org/10.1038/s41467-018-03695-x.
[83] Jason Horng, et al., Imaging electric field dynamics with graphene optoelectronics, Nat. Commun. 7 (2016) 13704. https://doi.org/10.1038/ncomms13704.
[84] Yuchen Zhang, et al., Probing electrical signals in the retina: via graphene-integrated microfluidic platforms, Nanoscale 8 (45) (2016) 19043–19049. https://doi.org/10.1039/c6nr07290a.
[85] Lixue Zou, et al., Trends analysis of graphene research and development, J. Data Inf. Sci. 3 (1) (2018) 82–100. https://doi.org/10.2478/jdis-2018-0005.
[86] By Gaurav Batra, Nick Santhanam, Kushan Surana, Graphene The Next S-Curve for Semiconductors?, 2018. https://www.mckinsey.com/industries/semiconductors/our-insights/graphene-the-next-s-curve-for-semiconductors.
[87] David Bradley, Is graphene safe? Mater. Today 15 (6) (2012) 230. https://doi.org/10.1016/S1369-7021(12)70101-3.
[88] Gaurav Lalwani, et al., Toxicology of graphene-based nanomaterials, Adv. Drug Deliv. Rev. 105 (2016) 109–144. https://doi.org/10.1016/j.addr.2016.04.028.
[89] Xiaoqing Guo, Nan Mei, Assessment of the toxic potential of graphene family nanomaterials, J. Food Drug Anal. 22 (1) (2014) 105–115. https://doi.org/10.1016/j.jfda.2014.01.009.
[90] Rickard Arvidsson, et al., Prospective life cycle assessment of graphene production by ultrasonication and chemical reduction, Environ. Sci. Technol. 48 (8) (2014) 4529–4536. https://doi.org/10.1021/es405338k.
[91] Rickard Arvidsson, Sverker Molander, Prospective life cycle assessment of epitaxial graphene production at different manufacturing scales and maturity, J. Ind. Ecol. 21 (5) (2017) 1153–1164. https://doi.org/10.1111/jiec.12526.
[92] P. Noorunnisa Khanam, et al., Biotechnological production process and life cycle assessment of graphene, J. Nanomater. (2017) 1–10. https://doi.org/10.1155/2017/5671584, 2017.
[93] Matteo Cossutta, Jon McKechnie, J. Stephen, Pickering, "A comparative LCA of different graphene production routes, Green Chem. 19 (24) (2017) 5874–5884. https://doi.org/10.1039/c7gc02444d.
[94] Michael Cai Wang, et al., A sustainable approach to large area transfer of graphene and recycling of the copper substrate, J. Mater. Chem. C 5 (43) (2017) 11226–11232. https://doi.org/10.1039/c7tc02487h.
[95] Ali Reza Kamali, Eco-friendly production of high quality low cost graphene and its application in lithium ion batteries, Green Chem. 18 (7) (2016) 1952–1964. https://doi.org/10.1039/c5gc02455b.
[96] Xiao-Ye Wang, Akimitsu Narita, Klaus Müllen, Precision synthesis versus bulk-scale fabrication of graphenes, Nat. Rev. Chem. 2 (December 2017). https://doi.org/10.1038/s41570-017-0100, 0100 (1–10).